\documentclass[10pt]{article}
\usepackage{amsmath}
\usepackage{graphicx}
\usepackage{amsfonts}
\usepackage{amssymb}
\usepackage{epsf}
\usepackage{latexsym}

\def\80{\hspace{0.8in}}

\newcommand{\be}{\begin{enumerate}}
\newcommand{\ee}{\end{enumerate}}
\newcommand{\bi}{\begin{itemize}}
\newcommand{\ei}{\end{itemize}}
\newcommand{\bd}{\begin{description}}
\newcommand{\ed}{\end{description}}

\def\beq{\begin{equation}}
\def\eeq{\end{equation}}
\def\bea{\begin{eqnarray}}
\def\eea{\end{eqnarray}}
\def\foo{\footnote}
\def\pa{\partial}
\def\d{\textrm{d}}

\def\mF{\mbox{F}}
\def\nF{\mbox{F}}

\def\mI{\mbox{I}}

\def\mM{\mbox{M}}

\def\sv{\mbox{\scriptsize v}}

\def\sA{\mbox{\scriptsize A}} 
\def\sB{\mbox{\scriptsize B}}

\def\sD{\mbox{\scriptsize D}}
\def\sE{\mbox{\scriptsize E}}
\def\sF{\mbox{\scriptsize F}}
\def\sG{\mbox{\scriptsize G}}
\def\sH{\mbox{\scriptsize H}}
\def\sI{\mbox{\scriptsize I}}

\def\sM{\mbox{\scriptsize M}} 
 
\def\sO{\mbox{\scriptsize O}}

\def\sR{\mbox{\scriptsize R}}
\def\sS{\mbox{\scriptsize S}}
\def\sT{\mbox{\scriptsize T}}

\def\sV{\mbox{\scriptsize V}}
\def\sW{\mbox{\scriptsize W}}

\def\eph(B){\mbox{\scriptsize emergent(LMB)}}

\def\fI{\mbox{\sffamily I}}

\def\fL{\mbox{\sffamily L}}

\def\fQ{\mbox{\sffamily Q}}

\def\fT{\mbox{\sffamily T}}
\def\fU{\mbox{\sffamily U}}
\def\fV{\mbox{\sffamily V}}

\begin{document}
\begin{titlepage}
\vspace{.7in}
\begin{center}

{\LARGE{\bf DOES RELATIONALISM ALONE}} 

\vspace{.1in}

{\LARGE {\bf CONTROL GEOMETRODYNAMICS WITH SOURCES?}} 

\vspace{.2in}

\large{\bf Edward Anderson}$^1$ 

\vspace{.2in}

\large{\em Peterhouse, Cambridge CB2 1RD}\normalsize

\vspace{.2in}

\large{and} \normalsize

\vspace{.2in}

\large{\em DAMTP, Centre for Mathematical Sciences, Wilberforce Road, Cambridge CB3 OWA.} \normalsize

\end{center}

\vspace{.2in}

\begin{abstract}

This paper concerns relational first principles from which the Dirac procedure exhaustively picks out 
the geometrodynamics corresponding to general relativity as one of a handful of consistent theories.  
This was accompanied by a number of results and conjectures about matter theories and general features 
of physics -- such as gauge theory, the universal light cone principle of special relativity and the 
equivalence principle -- being likewise picked out.  
I have previously shown that many of these matter results and conjectures are contingent on 
further unrelational simplicity assumptions.   
In this paper, I point out 1) that the exhaustive procedure in these cases with matter fields is slower than it was previously held to be.  
2) While the example of equivalence principle violating matter theory that I previously showed how to accommodate on 
relational premises has a number of pathological features, in this paper I point out that there is 
another closely related equivalence principle violating theory that also follows from those premises and is less 
pathological.  
This example being known as an `Einstein--aether theory', it also serves for 3) illustrating limitations 
on the conjectured emergence of the universal light cone special relativity principle.  

\end{abstract}

\vspace{.2in}

%

\vspace{.2in}


\vspace{1in} 

\noindent $^1$ ea212@cam.ac.uk

\end{titlepage}

\section{Introduction}

\subsection{Relationalism}

The relational perspective of Barbour \cite{BB82, Buckets, B94I, RWR, EOT} implements ideas of Leibniz 
\cite{Leibniz} and Mach \cite{Mach} (see also \cite{DOD}) to modern physics.    
In this approach, one starts with a configuration space $\fQ$ of (models of) whole-universe systems.  
One then adopts two relational postulates.

\noindent {\bf Configurational relationalism}: that certain transformations acting on $Q$ are   
physically meaningless.  
One way \cite{Lan}\footnote{Barbour's   
own way of conceptualing about configurational relationalism (`best 
matching'), see e.g. \cite{BB82, B03} is that, given two configurations, one should be kept fixed and 
the other should be shuffled around until an identification is found that minimizes its incongruence 
with the first one.  
The arbitrary frame method described in the main text here permits the form of the shuffling correction 
to be derived.    
Both approaches can be carried out for multiplier or velocity of a cyclic coordinate interpretations of 
auxiliaries in simple cases (which include all of those covered in this paper).}
of implementing this is to use arbitrary-$G$-frame-corrected quantities rather than bare $\fQ$ 
configurations, where $G$ is the group of physically meaningless motions.  
For, despite this augmenting $\fQ$ to $\fQ \times G$, variation with respect to each adjoined 
independent auxiliary $G$-variable produces a constraint which removes one $G$ variable and one 
redundancy among the $\fQ$ variables, so that one ends up on the quotient space $\fQ/G$ (the desired 
reduced configuration space).  
This is widely a necessity in theoretical physics through working on the various reduced spaces directly 
often being technically unmanageable, for instance in particle physics theories with its internal gauge 
group $G$ or in the split spacetime approach to general relativity with its spatial diffeomorphisms.

\noindent {\bf Temporal relationalism}: that there is no meaningful primary notion of time for the 
universe as a whole.  
One implementation of temporal relationalism is through using manifestly reparametrization invariant 
actions that do not rely on any extraneous time-related variables either.

For $\fQ$ = \{n particle postions\} and $G$ the Euclidean group of translations and rotations, the 
relational postulates form plausible alternative foundations for a portion of Newtonian mechanics 
\cite{ERPM, ERPMSRPM, POTERPM} (and admit also a scale-free counterpart for G the Similarity group of 
translations, rotations and dilatations \cite{B03, SRPM, ERPMSRPM}). 
The main idea in this paper concerns that (spatially compact without boundary) general relativity can be 
derived from these postulates in the case in which $G$ is the group of 3-diffeomorphisms. 
(This derivation also assumes a set of mathematical simplicity postulates and observational checks 
\cite{RWR, Phan} described in Sec 1.3).  
This answers a question of Wheeler: {\it ``if one did not know the Einstein--Hamilton--Jacobi equation, 
how might one hope to derive it straight off from plausible first principles without ever going through 
the formulation of the Einstein field equations themselves?" \normalfont (\cite{Battelle}, p 273) 
(Hojman, Kucha\v{r} and Teitelboim \cite{HKT} had previously provided a  distinct answer in which 
spacetime structure was presupposed; the present answer presupposes less structure than that, being 
a 3-space rather than split spacetime approach).  
Finally relational particle models have a number of useful analogue features permitting them to serve as useful 
\cite{K92, POTToys, POTERPM} toy model analogues for investigations of such as the problem of time 
in quantum gravity \cite{K92, I93}.

\subsection{General relativity admits a relational formulation}

One should first demonstrate that general relativity can indeed be recast as a 3-space approach theory.  
The Einstein--Hilbert action for the spacetime formulation of general relativity,\footnote{Here, 
$g_{AB}$ is the spacetime metric with determinant $g$ and Ricci scalar ${\cal R}$.
$h_{ab}$ is the induced 3-metric on a positive-definite 3-surface $\Sigma$ (interpreted, for the moment,  
as a spatial hypersurface within a spacetime), with determinant $h$, covariant derivative $D_a$ and Ricci 
scalar $R$.  
$\alpha$ is the lapse and $\beta_{\mu}$ is the shift. 
$\delta_{\beta} = \dot{\mbox{ }} - \pounds_{\beta}$ is the hypersurface derivative, where the dot is 
$\frac{\pa}{\pa\lambda}$ and $\pounds_{\beta}$ is the Lie derivative with respect to $\beta_a$.}     
\beq
\fI^{\sE\sH}_{\sG\sR}[g_{AB}] = \int\d^4x\sqrt{|g|}{\cal R} \mbox{ } ,
\eeq
when split with respect to a family of spatial hypersurfaces takes the conventional form \cite{ADM, 
DeWitt}
\beq
\fI^{\sA\sD\sM}_{\sG\sR}[h_{ab}, \alpha, \beta_a, \dot{h}_{ab}] = 
\int\d\lambda\int\d^3x\sqrt{h}\alpha\left\{\frac{\fT^{\sA\sD\sM}_{\sG\sR}}{4\alpha^2} + R \right\}  
\eeq
for
\beq
\fT^{\sA\sD\sM}_{\sG\sR} = \frac{1}{\sqrt{h}}G^{abcd}\{\delta_{\beta}h_{ab}\}\delta_{\beta}h_{cd}
\eeq
and
\beq
G^{abcd} = \sqrt{h}\{h^{ac}h^{bd} - h^{ab}h^{cd}\}  
\eeq
the inverse of the DeWitt supermetric \cite{DeWitt}.

A more useful prototype 3-space approach action \cite{BB82} can be formed by Baierlein, Sharp and 
Wheeler's \cite{BSW} procedure: solve the $\alpha$-multiplier equation 
$R-\fT^{\sA\sD\sM}_{\sG\sR}/4\alpha^2 = 0$ for $\alpha$ itself, 
$\alpha = \frac{1}{2} \sqrt{\fT^{\sA\sD\sM}_{\sG\sR}/R}$, and then use this to algebraically eliminate 
$\alpha$ from the Arnowitt--Deser--Misner Lagrangian.  
Thus one obtains 
\beq
\fI^{\sB\sS\sW}_{\sG\sR}[h_{ab}, \beta_a, \dot{h}_{ab}] = \int\d\lambda\int\d^3x\sqrt{h}
\sqrt{R\fT^{\sA\sD\sM}_{\sG\sR}} \mbox{ } .
\label{VBashwe}
\eeq
This is not quite reparametrization invariant because the shift is considered to be a coordinate for the 
purposes of variation.  
However, the Arnowitt--Deser--Misner split can be replaced by a split in terms of an instant variable 
$\mI$ (such that $\alpha = \dot{\mI}$) and a grid variable $\mF_a$ (such that $\beta_a = \dot{\mF}_a$, 
which is an example of a frame variable) at the pre-variational level if one takes into careful account 
that the auxiliary variables should be varied with free end spatial hypersurfaces 
\cite{ADMI}.\footnote{See  
\cite{B03, ABFO, ABFKO, ADMII} for earlier and further discussion of these variational methods.} 
%
Then one has an action 
\beq
\fI^{\sA}_{\sG\sR}[h_{ab}, \dot{h}_{ab}, \dot{\mF}_a, \dot{\mI}] = 
\int\d\lambda\int\d^3x\sqrt{h}\dot{\mI}\left\{\frac{\fT^{\sA}_{\sG\sR}}{4\dot{\mI}^2} + R \right\} 
\mbox{ } .  
\eeq
for 
\beq
\fT^{\sA}_{\sG\sR} = \frac{1}{\sqrt{h}}G^{abcd}\{\delta_{\dot{\sF}}\}h_{ab}\delta_{\dot{\sF}}h_{cd}
\mbox{ } .  
\eeq
Then performing Routhian reduction on this to eliminate $\dot{\mI}$ works out exactly the same as 
Baierlein--Sharp--Wheeler multiplier elimination, giving 
\beq
\fI^{\sA^{\prime}}_{\sG\sR}[h_{ab}, \dot{h}_{ab}, \dot{\nF}_a] = \int\d\lambda\int\d^3x\sqrt{h}
\sqrt{R\fT^{\sA}_{\sG\sR}} \mbox{ } .  
\eeq
This may now be taken as a starting point as done in \cite{RWR, ABFO, Lan} that implements the 
relational principles, in which case I use the notation $\&_{\dot{\sF}}$ for arbitrary G frame 
corrected derivative, here for $G$ the 3-diffeomorphisms on $\Sigma$: 
\beq
\fI^{\sT\sS\sA}_{\sG\sR}[h_{ab}, \dot{h}_{ab}, \dot{\mF}_a] = \int\d\lambda\int\d^3x \sqrt{h}
\sqrt{R\fT^{\sT\sS\sA}_{\sG\sR}[h_{ab}, \dot{h}_{ab}, \dot{\mF}_a]} 
\mbox{ } , \mbox{ } 
\fT_{\sG\sR}^{\sT\sS\sA} = \{h^{ac}h^{bd} - h^{ab}h^{cd}\}
\{\&_{\dot{\sF}}h_{ab}\}\&_{\dot{\sF}}h_{cd} \mbox{ } ,
\label{TSAGR}
\eeq
rather than the hypersurface derivative notation $\delta_{\dot{\sF}}$ that presupposes spacetime.  
Of course, in the present case, spacetime is nevertheless recovered.

\subsection{The `relativity without relativity' result}

Suppose next that one presupposes less structure: just 3-space notions rather than `3-space within 
spacetime' notions.    
Does general relativity then emerge?  
Does it emerge alone?  
One goes about investigating these questions using the {\it Dirac procedure} \cite{Dirac}.  
This involves taking the constraints that arise purely from the form of the Lagrangian without any 
variation (primary constraints) and those that have arisen so far by the variational process (secondary 
constraints), and demanding that these be propagated by the theory's evolution equations.  
This can lead to new constraints, in which case the Dirac procedure is applied again to these.  
Now, as each new constraint uses up some degrees of freedom (usually per space point in the present 
field-theoretic context) and the trial system has a finite amount of these, if the Dirac procedure runs 
through enough iterations, it uses up at least as many degrees of freedom as the trial theory had to 
start off with (see e.g. \cite{AB, Thiemann}.  
In this case, the trial theory has been demonstrated to be undesirable in being inconsistent (less than 
no degrees of freedom left), trivial (no degrees of freedom left) or undersized (e.g. a few global 
degrees of freedom alone could survive due to the shapes of the restrictions caused by the constraints, 
see e.g. \cite{Lan}).    
Then the only remaining way out is to restrict the trial theory by allowing some of the constraints to 
dictate how some of its hitherto free non-variational parameters should be fixed, and so one is 
exhaustively removing a number of the trial options.  
Thus the Dirac procedure lends itself to proofs by exhaustion.

By this method the `relativity without relativity' result arises: if one does not presuppose general 
relativity but rather to start with a wide class of reparametrization-invariant actions built out of 
good 3-d space objects in accord with the relational principles \cite{RWR, San, Van, Lan, Phan}, 
general relativity emerges.  
More concretely the input trial ans\"{a}tze are 
\beq
\fT_{\mbox{\scriptsize grav$($trial$)$}} = 
\frac{1}{\sqrt{h}Y}G^{abcd}(W)\{\&_{\dot{\sF}}h_{ab}\}\&_{\dot{\sF}}{h}_{cd} \mbox{ } ,
\label{VASBSW} 
\eeq  
for the gravitational kinetic term that is {\bf homogeneous quadratic in the velocities}, where
\beq
G^{ijkl}(W) \equiv \sqrt{h}  \{h^{ik}h^{jl} - Wh^{ij}h^{kl}\} \mbox{ } , \mbox{ }  
W \neq \frac{1}{3} \mbox{ } ,
\eeq
is the inverse of the most general ({\bf invertible}, {\bf ultralocal}) supermetric 

\noindent
\beq
G_{abcd}(X) = \frac{1}{\sqrt{h}} 
\left\{
h_{ac}h_{bd} - \frac{X}{2}h_{ab}h_{cd} 
\right\} 
\mbox{ } , \mbox{ }
X = \frac{2W}{3W - 1} \mbox{ } , 
\eeq  
and
\beq
\fV_{\mbox{\scriptsize grav$($trial$)$}} = A  + BR
\eeq
for the gravitational potential term.
This is {\bf second-order in spatial derivatives}.
The {\bf local square root} action is then
\beq
\fI_{\mbox{\scriptsize grav$($trial$)$\normalsize}}[h_{ab}, \dot{h}_{ab}, \dot{\mF}_i] = 
\int\d\lambda\int\d^3x \sqrt{h}\sqrt{\fV_{\mbox{\scriptsize grav$($trial$)$\normalsize}}
\fT_{\mbox{\scriptsize grav$($trial$)$\normalsize}}}
\mbox{ } .
\eeq
[I use bold font to denote what assumptions are being made; all of the assumptions in this 
Subsection are {\sl mathematical simplicity postulates} rather than deep physical principles.]

Then, setting $\dot{\mM}$ to be the emergent quantity $\frac{1}{2}\sqrt{
\fT_{\mbox{\scriptsize grav$($trial$)$}}^{\sT\sS\sA}/\fV_{\mbox{\scriptsize grav$($trial$)$}}}$,
the gravitational momenta are
\beq
\pi^{ab} \equiv \frac{\pa\fL}{\pa\dot{h}_{ab}} = 
\frac{\sqrt{h}Y}{2\dot{\mM}}G^{abcd}(W){\&}_{\dot{\sF}}{h}_{cd} \mbox{ } ,
\label{wmom}
\eeq
which are related by a primary constraint
\beq
{\cal H }_{\mbox{\scriptsize grav$($trial$)$}}\equiv YG_{abcd}(X)\pi^{ab}\pi^{cd}-\sqrt{h}\{A + BR\} = 0
\mbox{ } .
\label{VGRHam}
\eeq
Additionally, variation with respect to $\mF_a$ leads to a secondary constraint that is the usual
momentum constraint 
\beq
{\cal H}_{a} = D_b{\pi_a}^b = 0
\eeq 
thereby ensuring that the physical content of the theory is in the shape of the 3-geometry and not in 
the coordinate grid painted on it.
The propagation of ${\cal H}_{\mbox{\scriptsize grav$($trial$)$}}$then gives \cite{San, Lan}  
\beq
\dot{{\cal H}}_{\mbox{\scriptsize grav$($trial$)$}}
\approx \frac{2}{\dot{\mM}}\{X - 1\}BYD_i\{\dot{\mM}^2 D^i \pi\} \mbox{ } ,
\label{mastereq}
\eeq
[where $\approx$ is Dirac's notion of weak equality, i.e. equality up to (already-known) constraints].

From this, the main output is the `relativity without relativity' result that the Hamiltonian constraint propagates if the coefficient
in the supermetric takes the DeWitt value $X = 1 = W$.
In this case, embeddability of the 3-space into spacetime is recovered.
This in no way determines whether the emergent spacetime's signature is Lorentzian ($B=-1$) or Euclidean 
($B =1$): that is to be put in by hand.

\subsection{General relativity as geometrodynamics does not arise alone in the 3-space approach}

As it has 3 further factors \cite{RWR, San, Than, Phan}, (\ref{mastereq}) can vanish in 3 other ways.

\noindent 1) $B = 0$ gives strong or `Carrollian' gravity options regardless of whether $W = 1$ or not.  
The $W = 1$ case is the strong-coupled limit of general relativity \cite{SGLit}, which is a regime in 
which distinct points are causally disconnected by their null cones being squeezed into lines.  
This is relevant as an approximation to general relativity near singularities.
While, for $W \neq 1$, it is a similar limit of scalar--tensor theories \cite{San}.
In fact, all of these $B = 0$ options exist in two different forms: one without a momentum constraint
which thus are temporally but not spatially relational {\sl metrodynamics} and one with a momentum 
constraint which are are other consistent theories of geometrodynamics different to that obtained 
from decomposing the spacetime formulation of general relativity.

\noindent 2) $Y = 0$, gives `Galilean' theories.  
Here, the null cones are squashed into planes and there is no gravitational kinetic term.
Strictly speaking, for this option to arise, one should start with the Hamiltonian version of the
`Galilean' theory (as its degeneracy leads to there being no corresponding Lagrangian).

\noindent 3) $\pi = 0$ or $\pi/\sqrt{h}$ = constant preferred slicing conditions make the fourth factor 
vanish.  
This gives rise to alternative theories of conformal gravity \cite{ABFO} and to a {\sl derivation of general 
relativity, alongside the conformal method of treating its initial-value problem} \cite{York}{\sl, from a 
relational perspective} \cite{ABFO, ABFKO, ADMII}.   
These theories can be recast by restarting with an enlarged irrelevant group $G$ that consists of the 
3-diffeomorphisms together with some group of conformal transformations.

\subsection{Inclusion of matter in the 3-space approach}

The second theme of the 3-space approach papers concerns the inclusion of fundamental matter.
This is important for the 3-space approach, both as a robustness test for the axiomatization and to 
establish whether special relativity and the equivalence principle are emergent or require 
presupposition in this approach.

The robustness test is passed: using first constructive techniques \cite{RWR, AB} and then Kucha\v{r}'s  
\cite{K76} split spacetime framework techniques (\cite{Van}, see also Sec 3), all of minimally-coupled 
scalars, electromagnetism, Yang--Mills theory, Dirac theory, and all the associated gauge theories were 
found to be admitted by the 3-space approach.
There were some claims as regards well-known matter field types and physical principles being picked out.  
For example, it was claimed that 

\noindent 1) That electromagnetism and Yang--Mills theory are uniquely picked out \cite{RWR, AB}. 

\noindent 2) That minimally coupled scalars and 1-forms share null cones among themeselves (which is 
evidence toward the emergence of the special relativity principle).  
This is through each being forced to share the gravitational null cone.  

\noindent 3) That the equivalence principle is emergent rather than assumed.    

These were based on the simplicity assumptions of {\bf matter kinetic terms homogeneous quadratic in 
their velocities}, {\bf no metric-matter kinetic cross terms}, 
{\bf no matter dependence in the kinetic metric}.  

However, the split spacetime framework and allied techniques proved powerful enough to include massive 
(and other) vector fields \cite{Lan, Than, Phan} if these simplicity postulates are weakened in various 
ways, showing that the latter claims are partly based on mere simplicities that have nothing to do with 
relationalism.  
Hence 1) is false.  
Furthermore, this paper demonstrates that 2) and 3) are false.  
Essentially the split spacetime framework suggests further terms for the kinetic and potential ansatze 
with the inclusion of which further consistent theories can be cast in 3-space approach form.  
Overall, the relational postulates do not pick out the fields of nature, they include a wider range 
of fields.

A further new point I make in the present paper is that even within the simplicity postulates assumed, 
the claims were based on calculations that have two further tacit simplicity assumptions, 
without which the exhaustion rate would be slower than it was held to be.  

\noindent 1) {\bf Linear combination constraint preclusion.} In the original calculations, constraints 
arising as linear combinations of terms with different a priori free parameters were not considered 
to be a possibility.   
However, there is no good theoretical reason to preclude such constraints from arising.

\noindent 2) {\bf Second class constraint preclusion.} The original calculations' counting implicitly 
assumed that all constraints arising were first-class as regards how many degrees of freedom they used 
up.\footnote{A 
constraint is {\it first-class} if its Poisson brackets with all the other constraints close, and 
{\it second-class} otherwise.  
First-class constraints use up two degrees of freedom each while second-class ones use up just one.   
However one does not know before the Dirac process terminates whether a constraint is first or second  
class -- do its Poisson brackets with as yet unfound constraints from further along the Dirac 
progress close?  
Thus one cannot argue for emergent constraints to use up two degrees of freedom each (at least until 
the Dirac process has terminated and one has evaluated all those Poisson brackets).}  

\noindent While preclusion 2) is a brief and mathematically well defined simplicity postulate, it is 
highly restrictive, e.g. it does not cover the usual presentation of the phenomenologically useful 
massive vector field.  
This sort of restriction makes it very desirable from a theoretical perspective to uplift this simplicity.  
One way to proceed as regards 2) (which is simple and rigorous, although it is clearly not the most 
efficient) is to only assume that each constraint uses up at least one degree of freedom.

1) and 2) are clearly then capable of increasing the number of iterations required before a theory is 
shown to be inconsistent.      
In particular, for a set of interacting vector fields, weakening 1) costs one the capacity to produce 
internal index valued constraints at each step, meaning that one can no longer can one work for 
`arbitrary' gauge group.\footnote{`Arbitrary' 
here is subject to the (usual) requirement of being a direct sum of compact simple and U(1) Lie 
subalgebras so that the Gell-Mann--Glashow theorem applies \cite{Weinberg2}.}
%
All that is known now then is that for {\sl fairly small} gauge groups the calculation excludes 
alternatives, the calculation remaining unfinished for larger gauge groups. 
Thankfully, the gauge groups that have been found to explain experimental particle physics are not too 
large... 
%
%
(One can thus work furthermore case-by-case for larger groups required for more speculative 
theories of particle physics such as grand unified theories, while one should also not 
rule out that some new efficiency trick could be found so as to recover the result for an `arbitrary 
group').
One could likewise work case-by-case so as to safeguard other previous claims such as those about higher 
potential derivatives in vacuo in \cite{RWR}.

\subsection{Outline of the rest of this paper}

The constructive workings of \cite{RWR, AB}, all of which assume homogeneous quadratic kinetic terms 
with no metric--matter cross terms or matter field dependence in the kinetic metric, suffice as an  
arena in which to investigate the local emergence of special relativity (Sec 2), at least for simple 
3-space approach theories.  
In Sec 3, I recollect (and add to) arguments against the assertion (p 3217 of \cite{RWR})that in the 
3-space approach {\sl ``self-consistency requires that any 3-vector field must satisfy ... the 
equivalence principle"}. 
These arguments involve casting scalar(--vector)--tensor theories into 3-space approach form to act as 
counterexamples.  
I add further to these arguments in Sec 4 by constructing a unit vector tensor theory in 3-space 
approach form that is free of some pathologies common to vector--tensor theories and is both equivalence 
principle violating and special relativity violating in the sense that it has more than one distinct 
finite fundamental propagation speed.
Hence relationalism alone does not locally imply the special relativity principle.

\section{The position hitherto about the emergence of special relativity}

On p4 of \cite{Dirac}, Dirac explains that he uses actions so that relativity and gauge symmetry can be 
straightforwardly incorporated from the start.  
This is done by constructing one's action out of quantities that are Poincar\'{e} invariant for special 
relativity, diffeomophism--invariant for general relativity, U(1) gauge invariant for electromagnetic 
theory, and so on.
The 3-space approach is in a sense is a reverse of this: neither spacetime structure nor its locally 
special relativistic element are presupposed and it is shown that most alternatives to this are 
inconsistent.  
The way in which the early 3-space approach papers \cite{RWR, AB} include a range of standard bosonic 
matter fields minimally coupled to general relativity is a sufficient arena to investigate whether and 
how special relativity locally emerges in the 3-space approach.
These papers make the homogeneously-quadratic kinetic ansatz $\fT = \fT_{\Psi} + 
\fT_{\mbox{\scriptsize grav(trial)}}$, where the matter fields $\Psi_{\Delta}$ have kinetic term
\beq
\fT_{\Psi} = G^{\Gamma\Delta}(h_{ij})\{\&_{\dot{\sF}}\Psi_{\Gamma}\}\&_{\dot{\sF}}\Psi_{\Delta} \mbox{ },
\label{Kansatz}
\eeq
potential term denoted by $\fU_{\Psi}$ and momenta denoted by $\Pi^{\Delta}$.

Then the implementation of temporal relationalism by reparametrization invariance leads to a 
Hamiltonian-type constraint
\beq
{\cal H}_{\mbox{\scriptsize grav--$\Psi($trial$)$}} \equiv \sqrt{h}\{A + BR + \fU_{\Psi}\} - YG_{abcd}
(X)\pi^{ab}\pi^{cd} + \frac{G_{{\Gamma\Delta}}\Pi^{{\Gamma}}\Pi^{{\Delta}}}{\sqrt{h}} = 0 \mbox{ } .
\eeq
Applying Dirac's procedure and assuming that $\fU_{\Psi}$ at worst depends on connections (rather than 
their derivatives, which is true for the range of fields in question), the propagation of 
${\cal H}_{\mbox{\scriptsize grav,$\Psi($trial$)$}}$ gives
$$
\dot{{\cal H}}_{\mbox{\scriptsize grav--$\Psi($trial$)$}} \approx  \frac{2}{\dot{\mM}}D^a
\left\{
\dot{\mM}^2
\left\{
Y
\left\{
B
\left\{
D^b\pi_{ab} + \{X - 1\}D_a\pi
\right\}
+
\right.
\right.
\right.
\mbox{ } \mbox{ } \mbox{ } \mbox{ } \mbox{ } \mbox{ } \mbox{ } \mbox{ } \mbox{ } \mbox{ } \mbox{ }
\mbox{ } \mbox{ } \mbox{ } \mbox{ } \mbox{ } \mbox{ } \mbox{ } \mbox{ } \mbox{ } \mbox{ } \mbox{ }
\mbox{ } \mbox{ } \mbox{ } \mbox{ } \mbox{ } \mbox{ } \mbox{ } \mbox{ } \mbox{ } \mbox{ } \mbox{ }
$$
\beq
\left.
\left.
\left.
\left\{
\pi_{ij} - \frac{X}{2}\pi h_{ij}
\right\}
\left\{
\frac{\pa \mbox{\sffamily U\normalfont}_{\Psi}}{\pa {\Gamma^c}_{ia}}h^{cj} -
\frac{1}{2}\frac{\pa \mbox{\sffamily U\normalfont}_{\Psi}}{\pa{\Gamma^c}_{ij}}h^{ac}
\right\}
\right\}
+ G_{\Gamma\Delta}\Pi^{\Gamma}
\frac{\pa \mbox{\sffamily U\normalfont}_{\Psi}}{\pa(\pa_a\Psi_{\Delta})}
\right\}
\right\}
\mbox{ } ,
\label{sku}
\eeq
which is just an extension of (\ref{mastereq}) to include some matter fields.
The terms in (\ref{sku}) are then required to vanish for consistency.
This can occur according to various options, each of which imposes restrictions on
${\cal H}_{\mbox{\scriptsize grav--$\Psi($trial$)$}}$.
Furthermore, these options turn out to be very much connected to those encountered in the usual 
development of special relativity.

There is now a three-pronged fork in the choice of a universal transformation law in setting up special 
relativity.  
Two prongs are the Galilean or Lorentzian fork that Einstein faced (infinite or finite universal 
maximum propagation speed $c$).
The third prong is the Carrollian option $c = 0$.   
This last option occurs above through setting $B = 0$.
The vanishing of the other factors is attained by 1) declaring that $\fU_{\Psi}$ cannot contain 
connections.
2) It is then `natural' for $\mbox{\sffamily U}_{\Psi}$ not to depend on $\pa_a\Psi_{\Delta}$ either
(ultralocality in $\Psi_{\Delta}$), whereby the last term is removed.
Of course, we have good reasons to believe nature does not have $c = 0$, but what this option does lead
to is alternative dynamical theories of geometry to the usual general relativistic geometrodynamics.
Some are spatially relational and some are not.
This is an interesting fact from a broader perspective: it issues a challenge to why the 3-space 
approach insists on geometrodynamical theories since metrodynamical theories are also possible.
But what happens in the general relativity option is that the momentum constraint is an integrability 
condition \cite{MT72OM02, San} so one is stuck with geometrodynamics whether one likes it or not.

One could also enforce consistency above by the `Galilean' strategy of choosing  $Y = 0$.
This removes all but the last term.
It would seem natural to take this in combination with $\Pi^{\Delta} = 0$, whereupon the fields are not 
dynamical.
Moreover this does not completely trivialize the matter fields since they would then obey analogues of 
Poisson's law, or Amp\`{e}re's, and these are capable of governing a wide variety of complicated patterns.
Thus one arrives at an entirely nondynamical `Galilean' world.
In vacuo, this possibility cannot be obtained from a Baierlein--Sharp--Wheeler-type Lagrangian (the kinetic factor is badly
behaved) but the Hamiltonian description of the theory is unproblematic.
Of course, the Hamiltonian constraint is now no longer quadratic in the momenta:
\beq
{\cal H}_{\mbox{\scriptsize grav--$\Psi($trial$)$}}(Y = 0) = A + B R  + \fU_{\Psi} = 0 \mbox{ } .
\eeq
This option is not of interest if the objective is to find {\sl dynamical} theories.
Nevertheless, this option is a logical possibility, and serves to highlight how close parallels to the
options encountered in the development of special relativity arise within the 3-space approach.

There is also a combined locally Lorentzian physics and spacetime structure strategy as follows.
The signature is to be set by hand (one could just as well have any other nondegenerate signature
for the argument below).
Take (\ref{sku}) and introduce the concept of a gravity--matter momentum constraint 
${\cal H}_{        \mbox{\scriptsize grav--$\Psi$\scriptsize (trial)}         }^a$ by using 
$0 = -\frac{1}{2}{\cal H}_{\Psi}^a + \frac{1}{2}{\cal H}_{\Psi}^a$ and refactoring:
\noindent
$$
\dot{{\cal H}}_{\mbox{\scriptsize grav--$\Psi($trial$)$}}\mbox{$\approx$}\frac{2D^a}{\dot{\mM}}
\left\{
\dot{\mM}^2
      \left\{
Y
             \left\{
B
                   \left\{
                          \left\{
\underline{
D^b\pi_{ab} }
-
\underline{
\frac{1}{2}
\left\lfloor
\Pi^{\Delta}
\frac{\delta\pounds_{\dot{\mbox{\scriptsize F\normalsize}}}\Psi_{{\Delta}}}
{\delta\dot{\mF}^a}
\right\rfloor}
                          \right\}
+
\underline{
\frac{1}{2}
\left\lfloor
\Pi^{{\Delta}}
\frac{\delta\pounds_{\dot{\sF}}\Psi_{{\Delta}}}{\delta\dot{\mF}^a}
\right\rfloor}
                   \right)
            \right\}
+ \underline{G_{{\Gamma\Delta}}\Pi^{{\Delta}}
\frac{\pa \mbox{\sffamily U\normalfont}_{\Psi}}{\pa(\pa_a\Psi_{{\Delta}})}    }
      \right.
\right.
$$
\beq
\mbox{ } \mbox{ } \mbox{ } \mbox{ } \mbox{ } \mbox{ } \mbox{ } \mbox{ } \mbox{ } \mbox{ } \mbox{ } \mbox{ } \mbox{ } \mbox{ } \mbox{ } \mbox{ } \mbox{ } \mbox{ }
\left.
       \left.
            + \underline{YB\{X - 1\}D_a\pi}
            + \underline{Y
            \left\{
\pi_{ij} - \frac{X}{2}\pi h_{ij}
            \right\}
            \left\{
            \frac{\pa \mbox{\sffamily U\normalfont}_{\Psi}}{\pa {\Gamma^c}_{ia}}h^{cj}
          - \frac{1}{2}\frac{\pa \mbox{\sffamily U\normalfont}_{\Psi}}{\pa{\Gamma^c}_{ij}}h^{ac}
            \right\}    }
      \right\}
\right\}
\mbox{ } ,
\eeq
so that the first two underlined terms are then proportional to 
${\cal H}^a_{\mbox{\scriptsize grav--$\Psi($trial$)$}}$.\foo{
$\frac{\delta A}{\delta B}$ denotes the functional derivative, and the special brackets $\lfloor 
\mbox{ } \rfloor$ delineate the factors over which the implied integration by parts is applicable.}
%
In the `orthodox general covariance option', the third and fourth underlined terms cancel,
amounting to the enforcement of a universal null cone.
This requires supplementing by some means of discarding the fifth underlined term.
Here, one can furthermore {\sl choose} the orthodox option $X = 1$: the recovery of embeddability into 
spacetime corresponding to general relativity (the `relativity without relativity' result), or, 
{\sl choose} the alternative preferred-slicing worlds of $D_a\pi = 0$ which are governed by conformal 
mathematics.
As both of these options are valid, the recovery of locally-Lorentzian physics does not occur solely 
in generally-covariant theories.  
The connection terms (sixth underlined term) must also be discarded, but the Dirac procedure does 
this automatically for the given ans\"{a}tze.

Thus in the 3-space approach, locally-Lorentzian general relativistic spacetime arises as one option; 
other permitted options include Carrollian worlds, Galilean worlds and locally-Lorentzian preferred 
slicing worlds.
These alternatives all lack some of the features of generally relativistic spacetime.
\cite{Phan} went on to talk about hybrids of the above
The ultralocal and nondynamical strategies for dealing with the last term in (\ref{sku}) are available 
in {\sl all} the above options.
So as things stand, one does derive that gravitation enforces a {\sl unique finite} propagation speed,
but the possibility of coexisting with fields with infinite and zero propagation speeds is not precluded 
by consistency, although it does read to undersized solution spaces.\footnote{That 
such a dilemma exists was simply overlooked in \cite{RWR, AB} papers since it was claimed that these 
ultralocal and nondynamical strategies only lead to trivial theories, in the latter case by counting 
arguments.
Unfortunately, inspection of this triviality reveals it to mean `less complicated than in conventional
Lorentzian theories' rather than `devoid of mathematical solutions'.  
In particular, the counting argument is insufficient in not taking into account the geometry 
of the restrictions on the solution space.}
%
And of making the fourth underlined term vanish algebraically along the lines of parallel E and B in 
Poynting vector.
But none of these situations ruin the emergence of the special relativity principle in the sense that: 
any adjoined zero-momentum Galilean fields cannot propagate so that it does not matter that their 
propagation speed is in principle infinite, while adjoined Carrollian fields are precluded from 
propagating information away from any point by their ultralocal nature, and the parallel E and B field 
situation is also well-known to preclude the associated propagation (mutual orthogonality in the E and B 
fields causing each other to continue to oscillate).

However, we shall see in Sec 4 that the above fork breaks down for more complicated matter.

\section{The Position hitherto on the equivalence principle in the 3-space approach }

\subsection{Equivalence principle violations at the level of the action}

While this study started with partial evidence for the equivalence principle being emergent in the 
3-space approach \cite{RWR, Lan}, it then suffered the setback of counterexamples as more complete 
potential ansatze were devised.  
The counterexamples to date have, however, suffered from certain limitations.  
In this paper I extend the counterexamples to cases for which these limitations do not occur.      
I should first describe some symptoms at the level of action principles of whether a theory obeys or 
violates the equivalence principle.
Coordinates can be provided at each particular point $p$ such that the metric connection vanishes at $p$, 
so there is no obstruction in passage to the local special relativity form for curved spacetime matter 
field equations\footnote{N.B. 
that the gravitational field equations are given a special separate status in the equivalence principle
(`all the laws of physics bar gravity') and thus do not interfere with the logic of this.}
that contain no worse than metric connection.
However, the curvature tensor is an obstruction to such a passage if the field equations contain 
derivatives of the metric connection.  
Thus theories in which the matter terms contribute additional such terms are equivalence principle 
violating. 
One way in which derivatives of the metric connection in the field equations can arise from actions is 
through there already being such derivatives in the action e.g. in curvature--matter coupling terms.
A second way is from integration by parts during the variational working causing mere metric connections 
in the action to end up as derivatives of metric connections in the field equations.

\subsection{The split spacetime framework}

Rather than the previous sections' exhaustive Dirac procedure, this section requires the split 
spacetime framework, which does presuppose the general relativistic notion of spacetime.
The point of this is that there is then a systematic treatment of Kucha\v{r} \cite{K76} by which the 
spacetime formulation of specific consistent matter theories can be recast in split spacetime framework 
form.
It was using the split spacetime framework \cite{Van, Lan} that many matter theories were found to admit formulations that 
conform to the 3-space approach's relational principles (Sec 1.5).  
I next provide (as a new result) the variant of the split spacetime framework that is in terms of instant-grid variables 
for the case relevant here: a 1-form matter field.

One is presupposing that one has a hypersurface $\Sigma$ within a spacetime M.  
$n_{A}$ is the normal to $\Sigma$ and $e^a_A$ the projector onto this hypersurface.
Then it is meaningful to decompose each matter field into perpendicular and tangential parts with 
respect to $\Sigma$.  
In the case of the 1-form,
\beq 
A_A = n_A A_{\perp} + e^a_A A_a \mbox{ } .  
\label{trivia} 
\eeq 
Hypersurfaces can be re-gridded and deformed.  
Re-gridding kinematics involves Lie derivatives with respect to $\dot{\mF}_a$; these appear as 
corrections to the velocities so that these feature in the action as `hypersurface derivatives' 
rather than as `bare velocities'.  
As regards deformations, the arbitrary deformation of a hypersurface near a point $p$ splits into a 
{\it translation part} such that 
\beq
\dot{\mI}(p) \neq 0 \mbox{ } , \mbox{ } \mbox{ } \{\dot{\mI}_{,a}\}(p) = 0
\eeq 
and a 
{\it tilt} part such that 
\beq
\dot{\mI}(p) = 0 \mbox{ } , \mbox{ } \mbox{ } \{\dot{\mI}_{,a}\}(p) \neq 0 \mbox{ } .
\eeq
The translation piece further splits into a translation on a background spacetime piece and a 
{\it derivative coupling} piece which alters the nature of the background spacetime.
Furthermore, the re-gridding, tilt and derivative-coupling kinematics pieces that arise within this 
spacetime-presupposing framework are {\sl universal}: they depend solely on the rank of the tensor
matter field rather than on any details of that particular field.
It then so happens that two of these bear a tight relationship with what is needed to implement the 
configurational relationalism and temporal relationalism postulates.

\noindent 1) Use of the arbitrary 3-diffeomorphism frame is none other than re-gridding.

\noindent 2) The absence of tilt terms is a guarantee of an algebraic Routhian reduction procedure.
Thereby an extraneous time variable free action can be obtained, at least in principle.\foo{Were tilt 
terms not removable, as these contain spatial derivatives of $\dot{\sI}$, they would compromise the 
algebraicity of the elimination of $\dot{\sI}$ from the cyclic equation that arises from variation with 
respect to $\sI$.  
Of course, the algebraic equation might have no roots or only physically unacceptable (e.g. non-real) 
roots, in which cases the theory should be discarded.  
It might also not be explicitly soluble -- that is what I mean by `in principle'.}  
%
Tilts can however be removed from at least some actions by such as integration by parts or 
redefining field variables.

On the other hand, the derivative coupling universal feature is related to the equivalence principle, 
in that theories in which derivative coupling features in the split action are equivalence principle 
violating.  
Absense of derivative coupling is termed the {\bf geometrodynamical equivalence principle} in \cite{HKT}.  
It corresponds to the no metric--matter cross--term and no matter field dependence in the kinetic metric.
Thus these particular mathematical simplicity postulates additionally have physical significance.   
[But demanding that these hold amounts to imposing aspects of the equivalence principle by hand, 
so that one can no longer claim that the equivalence principle is emergent in the 3-space approach.]

This paper requires the split spacetime form of the derivatives of a 1-form, which is also 
a good illustration of re-gridding, tilt and derivative coupling terms (the last of these are 
those terms that involve the extrinsic curvature 
\beq
K_{ab} = -\frac{1}{2\dot{\mI}}\delta_{\dot{F}}h_{ab}
\eeq 
of the hypersurface).    
\beq
\nabla_bA_{\perp} = D_bA_{\perp} - K_{bc}A^c 
\mbox{ } ,
\label{Vderivproj1}
\eeq
\beq
\dot{\mI}\nabla_{\perp} A_{a} = - \delta_{\dot{\sF}}A_a - \dot{\mI} K_{ab}A^b - A_{\perp}\pa_a\dot{\mI}  
\mbox{ } ,
\label{Vderivproj2}
\eeq
\beq
\nabla_b A_a = D_bA_a - A_{\perp}K_{ab} 
\mbox{ } ,
\label{Vderivproj3}
\eeq
\beq
\dot{\mI}\nabla_{\perp}A_{\perp} = - \delta_{\dot{\sF}}A_{\perp} - A^a\pa_a\dot{\mI}  
\mbox{ } , 
\label{Vderivproj4}
\eeq
Intuitively, these relations come about because spacetime derivatives are not equal to spatial 
derivativess as the former have extra connection components, which this scheme interprets geometrically 
from the perspective of the hypersurface.

\subsection{Scalar--tensor theories}

The 3-space approach counterexample to date of the first type is Brans--Dicke theory \cite{BD} 
(see also \cite{KM} for a canonical treatment).    
While this was included in \cite{RWR} by casting the theory in the Einstein frame\footnote{Under 
this field redefinition, it is then a scalar field minimally coupled to gravity, which is clearly 
included among the 3-space approach castable cases listed in Sec 1.5.}
%
However, this transformation does away with the equivalence principle violation, so it is more 
instructive for the present context to work in Brans--Dicke theory's usual Jordan frame.  
For this, the spacetime action is 
\beq
\fI_{\sB\sD}[g_{AB}, \chi] = \int\d^4x\sqrt{|g|}\mbox{e}^{-\chi/2}\{{\cal R} - \omega\pa_A\chi\pa^A\chi\} \mbox{ } .
\eeq
The subsequent split spacetime action has a kinetic term proportional to 
\beq
\left\{
h^{ac}h^{bd} - \frac{X -2}{3X - 4}h^{ab}h^{cd}
\right\}
\delta_{\dot{\sF}}h_{ab}\delta_{\dot{\sF}}h_{cd}
+ \frac{4}{3X  -4}h^{ab}\delta_{\dot{\sF}}h_{ab}\delta_{\dot{\sF}}\chi 
+ \frac{3X - 2}{(3X - 4)(X - 1)}\delta_{\dot{\sF}}\chi\delta_{\dot{\sF}}\chi  
\mbox{ }   
\eeq
for 
\beq
X = \frac{2\{1 + \omega\}}{2\omega + 3} \mbox{ } .  
\eeq
Thus it is equivalence principle violating as it contains metric--matter kinetic cross terms.  
Nevertheless it can be cast into 3-space approach form \cite{Than} (but was missed in \cite{RWR} through 
the ansatze there not including metric--matter kinetic cross-terms).

However, this example  suffers the observational weakness that its parameter $\omega$ is fixed, expected 
on grounds of theoretical naturality to be of order unity and yet is bounded by the Cassini data to be above 20000 
\cite{BDBound}.  
This weakness can be removed by showing that the more general scalar--tensor theory with spacetime 
action\foo{This 
is not the most general scalar-tensor theory (see e.g. \cite{Will} and 
references therein). 
E.g. one could replace $e^{-\chi/2}$ by an arbitrary function of $\chi$, or 
furthermore extend the theory to have more than 1 scalar.  
However, the example in this paper is general enough to illustrate the point in question.} 
\beq
\fI_{\sS\sT\sT}[g_{AB}, \chi] = \int\d^4x\sqrt{|g|}e^{-\chi/2}\{{\cal R} - \omega(\chi)\pa_A\chi\pa^A\chi 
+ \fU(\chi)\}
\eeq
which one can likewise cast as a 3-space approach theory by performing the split with respect to a family of spatial 
hypersurfaces using instant--grid variables and then eliminating $\dot{\mI}$ and writing 
$\&_{\dot{\sF}}$ for $\delta_{\dot{\sF}}$ in the usual fashion.
That this can be cast in 3-space approach form is clear because adding a potential and replacing $\omega$ 
with $\omega(\chi)$ do not affect the split of the spacetime tensorial objects in the action or the form 
that the Routhian reduction that eliminates $\dot{I}$ are to take.  
While, this no longer suffers from the observational weakness because now $\omega$ varies and there is 
evidence that it tends dynamically to a large value in the late universe (toward the general relativity 
value of + $\infty$) \cite{GRatLateTimes}.

\subsection{Vector--tensor theories}

An example of 3-space approach theory \cite{Phan} in which the second kind of equivalence principle 
violation occurs can be found among the the vector--tensor theories considered in e.g. \cite{IN}. 
This class of theories has the spacetime form:
\beq
\fI_{\sV\sT\sT}[g_{AB}, A_A] = \int\d\lambda\int\d^3x\alpha
\left\{
{\cal R}   + \nu
\left\{
\nabla_A A^A\nabla_BA^B + m^2A^2
\right\}
\right\}
\mbox{ }
\label{NVA2}
\eeq
Now the split form of action (\ref{NVA2}) is (by the above derivative formulae and then using the field 
redefinition
\beq
\dot{\mI} A^a = \dot{v}
\label{light}
\eeq 
to remove `tilts' and also setting $A_{\perp}$ to be some $\phi$):
$$
\fI^{\mbox{\scriptsize ADM}}_{\sV\sT\sT}[h_{ab}, \dot{h}_{ab}, v_i, \dot{v}_i, \phi, \dot{\phi}, 
\dot{\sF}_i, \dot{\mI}] =
$$
\beq
\int\int \textrm{d}\lambda \dot{\mI}\sqrt{h}\textrm{d}^3x
\left\{
\frac{        \mbox{\sffamily T}^{\sA}_{\mbox{\scriptsize GR}}[h_{ab}, \dot{h}_{ab}, \dot{\mF}_i]         }
     {         4\dot{\mI}^2        } +
\frac{\nu}{\dot{\mI}^2}
\left\{
\left\{
D_a\{\dot{v}^a\} + \frac{\phi}{2}h^{ij}\delta_{\beta}h_{ij} + \delta_{\dot{\sF}}\phi
\right\}^2
+ {m^2\dot{v}^2}\right\} + R - \nu m^2\phi^2
\right\} \mbox{  } .
\eeq
Then a Routhian reduction of the same form as that mentioned in Sec 1.2 is possible, giving   
$$
\fI^{\sA^{\prime}}_{\sV\sT\sT}[h_{ab}, \dot{h}_{ab}, v_i, \dot{v}_i, \phi, \dot{\phi}, 
\dot{\mF}_i] 
=
$$
\beq
\int\int \textrm{d}\lambda \textrm{d}^3x\sqrt{h}
\sqrt{
\left\{
R - \nu m^2\phi^2
\right\}
\left\{
\fT^{\sA^{\prime}}_{\sG\sR} + 4\nu
\left\{\left\{
D_a\{\dot{v}^a\} + \frac{\phi}{2}h^{ij}\delta_{\dot{\sF}}h_{ij} + \delta_{\dot{\sF}}\phi
\right\}^2
+ m^2\dot{v}^2 \right\}
\right\}} \mbox{ } .
\eeq
[The equations encoded by this action happen to be weakly unaffected by whether $\dot{v}^a$ is replaced 
by $\delta_{\dot{\sF}}v^a$.]

Thus if one starts with 3-space approach principles, and using the arbitrary 3-diffeomorphism frame 
symbol $\&_{\dot{\sF}}$ in place of the hypersurface derivative symbol $\delta_{\dot{\sF}}$, one obtains 
the 3-space approach action
$$
\fI^{\mbox{\scriptsize TSA}}_{\sV\sT\sT}[h_{ab}, \dot{h}_{ab}, v_i, \dot{v}_i, \phi, \dot{\phi}, 
\dot{\sF}_i] =
$$
\beq
\int\int\d\lambda\d^3x\sqrt{h}\sqrt{
\left\{
R - \nu m^2\phi^2
\right\}
\left\{
\fT_{\sG\sR}[h_{ab}, \dot{h}_{ab}, \dot{\sF}_i] + 4\nu
\left\{
\left\{
D_a\{\&_{\dot{\sF}}v^a) + \frac{\phi}{2}h^{ij} \&_{\dot{\sF}}h_{ij} + \&_{\dot{\sF}}\phi
\right\}^2
+ m^2\{\&_{\dot{\sF}}v\}^2 
\right\}
\right\} } 
\mbox{ } .
\eeq
Thus one has a consistent (by reverse of above working and the original spacetime formulation being 
consistent) and nontrivial equivalence principle violating theory for geometry, a scalar and a 1-form.
It should be noted that \cite{RWR} missed this not on relational grounds but on simplicity grounds: 
the theory has a kinetic term that is not ultralocal, has metric--matter cross-terms and field 
dependence.

This was missed in \cite{RWR} through it having metric--matter kinetic cross-terms, matter field 
dependence in the kinetic metric and a mixture of 1-form and scalar modes from the 3-space perspective.

Many theories of this type have a number of undesirable features, such as classical and quantum 
instabilities \cite{IN, Clayton}, non-positivness of total energy \cite{J2} and formation of shocks 
beyond which the evolution cannot be extended \cite{Clayton}.  
\cite{Phan} speculated that some axiom that avoids such pathologies could be used to bring down this 
class of counterexample.

\section{A new example of 3-space approach theory that is all of equivalence principle violating, special relativity violating and less pathological}

\subsection{Unit vector--tensor theories (Einstein--Aether theories)}

\cite{J1,J2, Jacobson} consider a general Einstein--Aether action of the form 
\beq
\fI_{\sE\sA\sT}[g_{AB}, u_A] = \int \d^4x\sqrt{-g}\{{\cal R} + 
E_1\{\nabla_Au_B\}\nabla^Au^B + E_2\{\nabla_Au^A\}^2 + E_3\{\nabla_Au^B\}\nabla_Bu^A + 
E_4u^Au^B\{\nabla_Au_C\}\nabla_Bu^C + \lambda\{u_Au^A - 1\}\} \mbox{ } .
\eeq
As compared to the general theories considered by Isenberg and Nester, this permits 1 further derivative term 
(though I do not make use of it in my specific examples), and furthermore interprets what was the mass 
now as a Lagrange multiplier and adds the multiplier again as an extra potential piece.
[The Lagrange multiplier is there to implement the unit-field constraint.]    
At least some of these unit-field theories are less pathological \cite{J2}.

These theories are in general equivalence principle violators, the exception being if all of 
\beq
E_1 + E_3 = 0 \mbox{ } \mbox{ (Maxwellian combination) ,} 
\eeq
\beq
E_2 = 0 = E_4 \mbox{ } 
\eeq
hold.  
These theories are also in general special relativity violating, for they contain \cite{J1, J2} spin-2 
fields propagating at squared speeds 
\beq
c_{2} = \frac{1}{1 - \{E_1 + E_3\}} \mbox{ } ,
\eeq
spin-1 fields propagating at speed 
\beq
c_{1} = \frac{E_1 - E_1^2/2 + E_3^2/2}{\{E_1 + E_4\}\{1 - \{E_1 + E_3\}}
\eeq
and spin-0 fields propagating at speed 
\beq
c_{0} = \frac{\{E_1 + E_2 + E_3\}\{2 - \{E_1 + E_4\}\}}
{\{E_1 + E_4\}\{1 - \{E_1 + E_3\}\}\{2 + E_1 + E_3 + 3E_2 \}} \mbox{ } .  
\eeq
These are fairly extensively finite and with at least one distinct from the speed of light $c = 1$ 
in these units.  
This is the case unless all of  
\beq
E_1 + E_3 = 0 \mbox{ } , 
\eeq
\beq
E_4 = 0 \mbox{ } , 
\eeq
and 
\beq
E_1^{-1} - E_2^{-1} = 2 
\label{lumos}
\eeq
hold.

These squared speeds are also capable of going negative, corresponding to undesirable exponential-type 
instabilities.  
Positive linearized energy density requires 
\beq
\{2E_1 - E_1^2 + E_3^2\}/\{1 - E_1 - E_3\} > 0
\eeq
(vector mode contribution) and
\beq
\{E_1 + E_4\}\{2 - E_1 - E_4\} > 0
\eeq
(trace mode contribution).  
One would also like the kinetic energy contributions to have the usual sign for matter kinetic terms.

\subsection{Einstein--Aether theories that are castable in 3-space approach form}

To build a suitable 3-space approach example that is equivalence principle violating, special relativity violating in the sense of having 2 different 
finite fundamental propagation speeds and not subject to the above three pathologies, proceed as follows.

Consider first the theory with $E_2$ alone nonzero.   
Compared to the previous section's theory, the only difference is to the potential (which is trivial to 
split spacetime framework decompose), so the previous section's working will straightforwardly extend to   
the Einstein--aether theory case that is analogous to the above specially-chosen case. 
But for $E_2$ theory $c_2 = 1 ( = c)$ and the other two are not finite, so this does not constitute 
a special relativity violation of the type I am seeking.

But consider then furthermore including a Maxwell-type combination ($E_1 = -E_3 \neq 0$) in the action; 
as this has a very simple split spacetime framework  decomposition, this addition does not ruin the 
algebraicity of the Routhian reduction. 
So the theory I choose to work with is identified in the spacetime picture as 
\beq
\fI_{\sE\sA\sT}[g_{AB}, u_A] = \int \d^4x\sqrt{-g}\left\{{\cal R} + E_1\{\nabla_Au_B\}\nabla^Au^B + 
E_2\{\nabla_Au^A\}^2 - E_1\{\nabla_Au^B\}\nabla_Bu^A + \lambda\{u_Au^A - 1\}\right\} \mbox{ } .  
\eeq
split spacetime framework  splitting this, adhering to the redefinition (\ref{light}), using 
symmetry-antisymmetry cancellations on the new quadratic tilt terms and integration by parts on the new 
linear tilt terms, one indeed passes to a homogeneous quadratic action to which the usual Routhian 
reduction move can be carried out.  
The resulting action may, moreover be interpreted as (rewriting $\delta_{\dot{\sF}}$ as $\&_{\dot{\sF}}$ 
and adopting this action as one's new starting-point) a 3-space approach theory that follows from the 
configurational and temporal relationalism principles: 
\beq
\fI_{\sE\sA\sT}^{\sT\sS\sA}[h_{ab}, \dot{h}_{ab}, \dot{\mF}_a, \dot{v}_a] = 
\int\d\lambda\int\d^3x\sqrt{h}\sqrt{\fT\fU} 
\eeq
for
\beq
\fU = R - \mu\{\phi^2 + 1\}
\eeq
and
\beq
\fT = \fT^{\sB\sF\sO{-}\sA}_{\sG\sR} + \fT^{\sB\sF\sO{-}\sA}_{\sv}(\nu \rightarrow E_2, m^2 \rightarrow 
\mu) + \fT^{\sB\sF\sO{-}\sA^{\prime}}_{\sv} \mbox{ } , 
\eeq
where
\beq
\fT^{\sB\sF\sO{-}\sA^{\prime}}_{\sv} = E_1
\left\{       
\{  h^{ac}h^{bd} - h^{ad}h^{bc} \}
\pa_a\{    \&_{\dot{\sF}}     v_b     \}\pa_c      \&_{\dot{\sF}}     v_d + 
2D^b
\left\{     
\&_{\dot{\sF}}     v^a \{    \pa_b\&_{\dot{\sF}}     v_a  -  \pa_a\&_{\dot{\sF}}     v_b  \}   
\right\}  
+ 
\{         \&_{\dot{\sF}}      \phi    \}^2  
\right\} 
\mbox{ } .  
\eeq
Compared to the original example I gave, this is more general in having $E_2$ and less general in being 
a unit vector field.

Now, indeed, there is in general more than 1 fundamental propagation speed, as 
\beq
c_{2} = c_{1} = 1 \mbox{ } ( \mbox{ } = c) \mbox{ } , \mbox{ } c_0 = \frac{E_2\{2 - E_1\}}{E_1\{2 + 3E_2\}} \mbox{ } .
\eeq
I.e. this example contains 1) a non-generic case 
\beq
E_1^{-1} - E_2^{-1} = 2 
\eeq
which is {\sl not} a counterexample to violation of the special relativity lightcone (this is a subcase 
of the above non-special relativity violating example).    
2) The general case 
\beq
E_1^{-1} - E_2^{-1} \neq 2 
\eeq
for which 
\beq
c_0 \neq 1 = c_1 = c_2 = c \mbox{ } .  
\eeq
So there is a 1-parameter family (bar a single parameter value) of 3-space approach complying special relativity 
violating theories: there are scalar modes whose propagation speed in vacuo is different from the speed 
of light, so these have a null cone structure that is {\sl not} shared with the other fields in this theory.

A fair portion of the above example's parameter space is able to comply with positive linearized energy: 
that for which
\beq
0 < E_1 < 2 \mbox{ } .  
\eeq
For $E_2 > 0$ or $< -2/3$, this remaining region complies with the stability criteria $c_0^2$, $c_1^2$, 
$c_2^2 > 0$ since these reduce to the trivial $1 > 0$ (twice) and 
\beq
\frac{E_2\{2 - E_1\}}{E_1\{2 + 3E_2\}} > 0\mbox{ } ,   
\eeq
however the latter subregion should be discarded to ensure that the kinetic term is of the right 
characteristic sign for a matter contribution, thus leaving one with the `region of non-pathology' 
\beq
0 < E_1 < 2 \mbox{ } , \mbox{ } E_2 > 0 \mbox{ }   
\eeq
for the theory's coupling constants.

This region is split into two pieces by the curve of special relativity-compliance [i.e. of universal 
luminal fundamental speed (\ref{lumos})]; the subregion above this curve has the scalar modes propagate 
superluminally and the subregion below this curve has them propagate subluminally.  


\section{Conclusion}

The 3-space approach is based on temporal and configurational relational principles.  
General relativity in geometrodynamical form can be derived as one consistent alternative that follows 
from these premises when applied to a theory for which the 3-metrics on 
a fixed spatial topology are redundant dynamical objects under the associated 3-diffeomorphisms. 
A sufficient set of fundamental matter fields to describe nature can be adjoined to this scheme.  
It was furthermore claimed that 

\noindent 1) working with matter fields alongside spatial 3-metrics {\it picks out} 
electromagnetism (and Yang--Mills theory) coupled to general relativity as the only consistent theories 
of one (and K interacting) 1-forms.

\noindent 2) The equivalence principle is emergent.

\noindent 3) The universal null cone of special relativity is locally recovered.  

\noindent These were always subject to simplicity assumptions as well as the relational postulates.  
Claim 1) should be weakened, at least on basis of current workings.  
This is not only because lifting unrelational simplicities that were identified as such at the time of 
doing the calculation has been shown to destroy the result, but also because of two further tacit 
simplicities assumed in the proof, one of which is unmotivated and the other of which is unduly 
restrictive from a theoretical perspective.  
Without these the exhaustion goes more slowly and one then has to work case by case rather than once and 
for all with an arbitrary gauge group.  

Also, examples including some new to this paper show that it is necessary for the relational postulates 
to be supplemented by non-relational simplicity assumptions in order for 2) and 3) to hold.  
As these necessary simplicity assumptions include what Hojman, Kucha\v{r} and Teitelboim 
identify as the geometrodynamical equivalence principle, the 3-space approach's claim of 
{\sl deriving} the equivalence principle loses its credibility.   
Furthermore, from the split spacetime framework perspective, the geometrodynamical equivalence principle 
and statements equivalent to the relational postulates come as a neat package involving the three types 
of universal kinematics: hypersurface derivatives, the absense of tilts and the absense of 
derivative couplings, while taking the 3-space approach and the geometrodyamical equivalence principle 
as one's principles is more heterogeneous.\foo{The 3-space approach 
assumes less structure than (split) spacetime approaches.  
That makes it `more interesting' but also harder to work with as there being less structure 
makes proving thoerems harder in the 3-space approach than e.g. in Hojman--Kuchar--Teitelboim's 
approach that presupposes spacetime.  
E.g., their use of induction proofs specifically rely on additional spacetime structure.}
%
That said, the equivalence principle is separate from other postulates in Einstein-type spacetime approach, so one is doing 
no worse than what one does in taking the geometrodynamical equivalence principle alongside the 
relational postulates to be the heart of the axiomatization of general relativity.  
That reflects the primality of the equivalence principle as regards axiomatizations of general relativity -- so far as the author (or 
Brown \cite{HB}) are aware, no derivations of the equivalence principle from more basic postulates are known (which 
is what merited my concentrated effort to bring down \cite{RWR}'s conjecture otherwise).

`Simple' (in the sense of Sec 1.5) matter fields coupled to dynamical 3-metrics builds in the 
equivalence principle; one then encounters (an extension of) the fork Einstein encountered in setting up 
special relativity as the ``roots of" an explicit equation arising from the Hamiltonian-type constraint 
by the Dirac procedure.  
These correspond to Lorentzian relativity (single finite physical propagation speed), 
Galilean relativity (infinite propagation speed), and Carrollian relativity (zero propagation speed).
But if the associated simplicities are dropped, this article's example shows that equivalence principle violation is possible including in 
otherwise relatively non-pathological situations, and more than 1 finite fundamental propagation speed 
can occur -- consistency other than by the above fork becomes allowed.

A new issue to investigate -- I dare not call it a conjecture -- is whether each of the local recovery 
of special relativity and of gauge theory can be shown to follow from the equivalence principle free of 
the 3-space approach or even geometrodynamical formalism.  
Here, `shown to follow' might mean that they are among the natural structures to emerge, and perhaps a 
further axiom or a collection of observations or demands from local quantum field theory could remove 
some (or all) of the structures that co-emerge with them.  
This does not look to be restricted to a specific geometrodynamical formulation or even to 
geometrodynamics, but would rather be a stronger formalism-independent result of general relativity. 

\mbox{ }

\noindent{\bf Acknowledgments}

\mbox{ }

\noindent I thank Julian Barbour and Harvey Brown for previous discussions and Peterhouse Cambridge for 
funding me in 2007-2008.  


\end{document}